\documentclass{amsart}

\usepackage{float}
\restylefloat{table}
\usepackage[utf8]{inputenc}
\usepackage[toc,page]{appendix}
\usepackage{amsmath,amsthm,amssymb,stmaryrd}
\usepackage{graphicx}
\usepackage[colorinlistoftodos,prependcaption]{todonotes}
\usepackage{xargs}
\usepackage{multicol}
\newcommandx{\unsure}[2][1=]{\todo[linecolor=blue,backgroundcolor=blue!25,bordercolor=blue,#1]{#2}}
\newmuskip\pFqmuskip

\newtheorem{theorem}{Theorem}[section]

\newcommand*\pFq[6][8]{%
	\begingroup 
	\pFqmuskip=#1mu\relax
	\mathcode`\.=\string"8000
	\begingroup\lccode`\~=`\,
	\lowercase{\endgroup\let~}\pFqcomma
	{}_{\,#2}F_{\,#3}{\left[\genfrac..{0pt}{}{\,#4}{\,#5};#6\right]}%
	\endgroup
}
\newcommand{\pFqcomma}{\mskip\pFqmuskip}

\newcommand{\bbint}[2]{\ensuremath{\;\backslash\!\!\!\!\backslash\!\!\!\!\!\int_{#1}^{#2}}}

\begin{document}
\title[Generalized Stieltjes transform]{Finite-Part Integration of the Generalized Stieltjes Transform and its dominant asymptotic behavior for small values of the parameter. II. Non-integer orders}

\author{Christian D. Tica}
\author{Eric A. Galapon}
\address{Theoretical Physics Group, National Institute of Physics, University of the Philippines, Diliman Quezon City, 1101 Philippines}
\email{eagalapon@up.edu.ph}
\date{\today}

\maketitle
\begin{abstract}
The paper constitutes the second part on the subject of finite part integration of the generalized Stieltjes transform $S_{\lambda}[f]=\int_0^{\infty} f(x) (\omega+x)^{-\lambda}\mathrm{d}x$ about $\omega = 0$ where now $\lambda$ is a non-integer positive real number. Divergent integrals with singularities at the origin are induced by writing $(\omega+x)^{-\lambda}$ as a binomial expansion about $\omega = 0$ and interchanging the order of operations of integration and summation. The prescription of finite part integration is then implemented by interpreting these divergent integrals as finite part integrals which are  rigorously represented as complex contour integrals. The same contour is then used to express $S_{\lambda}[f]$ itself as a complex contour integral.  This led to the recovery of the terms missed by naive term-wise integration which themselves are finite parts of divergent integrals whose singularity is at the finite upper limit of integration. When the function $f(x)$ has a zero at the origin of order $m=0,1,\dots$ such that $m-\lambda<0$ the correction terms missed out by naive term by term integration give the dominant contribution to $S_{\lambda}[f]$ as $\omega\to 0$. Otherwise, the correction term is sub-dominant to the leading convergent terms in the naive term by term integration. We apply these results by obtaining exact and asymptotic representations of the Kummer and Gauss hypergeometric functions by evaluating their known Stieltjes integral representations. We then apply the method of finite part integration to obtain the asymptotic behavior of a generalization of the Stieltjes integral which is relevant in the calculation of the effective index of refraction of a shallow potential well.
\end{abstract}

\section{Introduction}
In this paper, we consider the evaluation of the incomplete generalized Stieltjes transform for non-integral order $\lambda$,
\begin{equation}\label{ogiy}
S_{\lambda}^{a}[f]=\int_0^{a}\frac{f(x)}{(\omega+x)^{\lambda}}\, \mathrm{d}x, \;\;\; a,\omega,\lambda>0,\;\; \lambda \neq 1, 2, 3, \dots ,
\end{equation}
about $\omega =0$ by finite part integration. This procedure is first introduced in \cite{galapon2} and applied in \cite{tica} to obtain both an exact and asymptotic representations of the generalized Stieltjes transform of integral orders for small values of the real positive parameter $\omega$. We will do the same to evaluate \eqref{ogiy} and take the limit $a\to\infty$ to obtain the corresponding result for the generalized Stieltjes transform of non-integer order\cite{widder,saxena,joshi,yurekli,karp}. The need for a separate treatment stems from the fact that the nature of the singularity of the complex valued function $(\omega + z)^{-\lambda}$ depends on $\lambda$: when $\lambda$ is an integer $n$, the integrand  in \eqref{ogiy} has a pole at $z=-\omega$ of order $n$; on the other hand, when $\lambda$ is non-integer, the integrand has a branch point at $z=-\omega$ instead. Hence, for the present case, neither Lemma 2.1 nor 2.2 of \cite{tica} gives the appropriate complex contour integral formulation of equation \eqref{ogiy}. Instead, we shall faithfully adhere to the sequence of steps prescribed in \cite{tica} to properly implement finite part integration and obtain the contour integral representation of the integral in equation \eqref{ogiy}. 

The first step is to take up the integration over the positive real line into the complex plane. This migration is facilitated by a contour suggested by the finite part of the divergent integral which we will now induce by the following expedient manipulation. Expanding the kernel $(\omega+x)^{-\lambda}$ about $\omega=0$ and then integrating term by term yields an infinite series of divergent integrals
\begin{equation}\label{soty}
\sum_{j=0}^{\infty} {-\lambda \choose j} \omega^{j} \int_0^{a} \frac{f(x)}{x^{ j +  \lambda}}\mathrm{d}x .
\end{equation}
The divergence occurs at the origin when $f(0)\neq 0$. Their corresponding finite parts have already been previously considered in \cite{galapon2,tica}. In particular, their contour integral representation is given by \cite[Theorem 2.2]{tica},
\begin{equation}
	\bbint{0}{a}\,\frac{f\left(x\right)}{x^{j+\lambda}}\,\mathrm{d}x=\frac{1}{\left(e^{-2\,\pi\,\lambda\,i}-1\right)}\,\int_{\mathrm{C}}\,\frac{f\left(z\right)}{z^{j+\lambda}}\,\mathrm{d}z,\qquad j=1,2,\dots, \lambda\neq 1,2,\dots
\end{equation}
The contour $\mathrm{C}$ is shown in Figure-\ref{around_-omega}. This is the same contour with which \eqref{ogiy} is written as the following complex contour integral,
\begin{equation}
\int_{\mathrm{C}}\frac{f(z)}{\left(\omega+{z}\right)^{\lambda}}\mathrm{d}z.
\end{equation}
This concludes the first step.

The precise description of the succeeding series of steps will be the content of Section-\ref{nonintegral}. In essence, this entails deforming $\mathrm{C}$ into an equivalent contour $\mathrm{C'}$ and by asserting the equivalence of integrating over these two contours, a complex contour integral representation for \eqref{ogiy} will emerge. It will be shown to take the form
\begin{equation}\label{soti}
	\int_0^a\frac{f(x)}{(\omega+x)^{\lambda}}\mathrm{d}x=\sum_{j=0}^{\infty}{{-\lambda}\choose {j}}\, {\omega}^j \, \bbint{0}{a}\frac{f(x)}{x^{j+\lambda}}\mathrm{d}x + \Delta_{\mathrm{sc}}^{\lambda}(\omega),
\end{equation}
where the first term in the right hand side is the summation in \eqref{soty} with the divergent integrals interpreted as their corresponding finite parts. The second term is the relevant correction or compensation term  to the rather arbitrary assignment of regularized value, finite part in our case, to the divergent integral. It is interesting to note that had we opted for a different regularization of the divergent integrals in \eqref{soty}, the nature of this compensation term will change correspondingly and will be determined by the set of rules or ``calculus" by which one gives rigorous representations to the specific regularized value we chose \cite{calc}. For the case of the Hadamard finite part, with the finite parts defined as complex contour integrals, the rules of finite part integration is precisely its corresponding calculus. This, in essence, is the thesis of a recent paper on the proper use of divergent integrals in calculation of the exact values of convergent integrals \cite{calc}. 

In \cite{tica}, the first term is called the naive term since it emerges from the naive interchange of the summation and integration. The second term is called the singular term as it arises from the contribution of the singularities encountered in the complex contour integration of equation \eqref{ogiy}. We shall use these nomenclature here for consistency. We shall see in Section-\ref{nonintegral} that the singular term will take the form of a finite part of a divergent integral with an end-point non-integrable singularity. We will  demonstrate in Section-\ref{endpoint} and in the succeeding examples how to extract the corresponding finite part both from the canonical definition and by equivalently relating it to the finite part corresponding to the divergent integral which results from confining the singularity at the origin by a trivial change of variable. These procedures will yield the same unique representations for the finite part which will then allow  us to promptly identify them as they emerge in the course of evaluating \eqref{ogiy} by finite part integration in Section-\ref{nonintegral}.

The rest of the paper is organized as follows. In Section-\ref{param}, we consider the asymptotic behavior of the generalized Stieltjes transform \eqref{ogiy} for $0<\omega\ll 1$ by determining the dominant term when the function $f(x)$ has a zero at the origin of some integer order that is either greater or less than the order $\lambda$ of the transform. In Section \ref{ung}, the result of Section-\ref{nonintegral} is applied to obtain an expansion for the Gauss hypergeometric function and the Kummer function of the second kind about $\omega=0$ by finite part integration of their integral representation written as generalized Stieltjes transform of non-integral order. In Section-\ref{appli}, we obtain both an exact and a suitable asymptotic expansion of the integral taking the form
\begin{equation}
\int_{0}^{\infty}\frac{f(x)}{\sqrt{\omega^2+x^2}}\mathrm{d}x
\end{equation}
for $0<\omega \ll 1$ and $f(x)$ possessing an entire complex extension $f(z)$. The result obtained here will be useful in obtaining the asymptotic behavior of the effective index of refraction \cite{galaponprl} of a finite one-dimensional quantum well \cite{alvin}. The $\omega\ll 1$ case corresponds to a physical situation where the potential well has a shallow depth. 
Finally, conclusions and motivations for further applications and investigations are briefly discussed in the Section-\ref{conclusion}. All analytical results are confirmed numerically using Mathematica\textsuperscript \textregistered  11.2 and Maple\textsuperscript \textregistered 18.00 on an Intel\textsuperscript \textregistered Core i7 processor with 8Gb of RAM.

\section{Finite part integral for an end-point non-integrable singularity}\label{endpoint}
When finite part integration of the incomplete generalized Stieltjes tranform \eqref{ogiy} lifts the integration from the positive real line into the complex plane as a complex  contour integral, the point $z=-\omega$ is no longer a pole of order $n$ which was the case with integer-ordered Stieltjes transform \cite{tica}, but  a branch point. Consequently, this will prompt us to handle divergent integrals of the form
\begin{equation}\label{another}
\int_0^c\frac{g(x)}{(c-x)^{n+\alpha}} \mathrm{d}x, \qquad 0<\alpha<1,\,\,n=1,2, \dots
\end{equation}
whose divergence arise from a non-integrable singularity at the finite upper limit of integration, $x=c$ . Incidentally, in the process of obtaining the finite part of \eqref{another}
, it is possible to perform a simple change of variable  $c-x \to x$ so as to confine the singularity at the origin. Since the rules governing such a step are well established when working with convergent integrals, it must be performed after the temporary removal of the singular point which  renders the divergent integral well-defined. 

This will then allow us to relate this finite part to the ones considered in the previous work, that is
\begin{equation}\label{kilo}
    \int_{0}^{c-\epsilon}\frac{g(x)}{\left(c-x\right)^{n+\alpha}}\mathrm{d}x = \int_{\epsilon}^{c}\frac{g(c-x)}{x^{n+\alpha}}\mathrm{d}x.
\end{equation}
Hence, upon taking the limit as  $\epsilon\to 0$, we see that the following finite parts are equal
\begin{equation}\label{miyo}
    \bbint{0}{c}\frac{g(x)}{\left(c-x\right)^{n+\alpha}}\mathrm{d}x = \bbint{0}{c}\frac{g(c-x)}{x^{n+\alpha}}\mathrm{d}x.
\end{equation}
The finite part integral in the right hand side of the equation above is given in Theorem 3.2 in \cite{tica}
\begin{align}
    \bbint{0}{c}\frac{g(c-x)}{x^{n+\alpha}}\mathrm{d}x 
    = \sum_{j=0}^{\infty}\frac{g^{(j)}(c)(-1)^j}{j!\,\left(j+1-n-\alpha\right)}c^{j+1-n-\alpha}
\end{align}
with the assumption that $g(c-x)$ is infinitely differentiable. The corresponding contour integral representation is again given in \cite[Theorem 2.2]{tica},
\begin{equation}\label{oit}
    \bbint{0}{c}\frac{g(c-x)}{x^{n+\alpha}}\mathrm{d}x = \frac{1}{e^{-2\pi\alpha i}-1}\int_{\mathrm{C}}\frac{g(c-z)}{z^{n+\alpha}}\mathrm{d}z,
\end{equation}
Where $\mathrm{C}$ is given in Figure-\ref{around_-omega}. 

It is possible to verify that equation \eqref{miyo} holds
by independently obtaining the same value for the finite part of the divergent integral \eqref{another} using the canonical definition given in \cite{galapon2,tica,monegato}. This entails expanding $g(x)$ about $x=c$ in the left hand side of equation \eqref{kilo} followed by interchanging the order of integration and summation. Taking the limit as $\epsilon\to 0$ and identifying the convergent and the divergent terms, it can be shown that the finite part in the left hand side of \eqref{miyo} indeed admits the following explicit representation from the convergent terms
\begin{equation}\label{nit}
    \bbint{0}{c}\frac{g(x)}{\left(c-x\right)^{n+\alpha}}\mathrm{d}x = \sum_{j=0}^{\infty}\frac{g^{(j)}(c)(-1)^j}{j!\,\left(j+1-n-\alpha\right)}c^{j+1-n-\alpha}
\end{equation}
and from the divergent group of terms
\begin{eqnarray}\nonumber\label{asol}
    \bbint{0}{c}\frac{g(x)}{\left(c-x\right)^{n+\alpha}}\mathrm{d}x = \lim_{\epsilon\to 0}\left[\int_{0}^{c-\epsilon}\frac{g(x)}{\left(c-x\right)^{n+\alpha}}\mathrm{d}x\right. \\
    \left. - \sum_{j=0}^{n-1}\frac{g^{(j)}(c)(-1)^j}{j!\left(j+1-n-\alpha\right)}\frac{1}{\epsilon^{n+\alpha-j-1}}\right].
\end{eqnarray}

Furthermore, by proceeding in a similar fashion, it is also possible to show that equation \eqref{miyo} remains true even for the case where $\alpha = 0$. That is, for all functions $g(x)$ with properties stated here, it is always true that
\begin{equation}
        \bbint{0}{c}\frac{g(x)}{\left(c-x\right)^{\lambda}}\mathrm{d}x = \bbint{0}{c}\frac{g(c-x)}{x^{\lambda}}\mathrm{d}x,
\end{equation}
for all real $\lambda\geq 1$. Contrary to the particular example considered in \cite[p9]{tica} which demonstrates how a linear change of variable directly applied on finite part integrals does not yield the appropriate result when $\lambda$ is a positive integer, the present case shows that such a step is generally meaningful when instead implemented on the convergent integral which appears in the interim step in the process of extracting the finite part. 

\section{Generalized Stieltjes transform of non-integral order of 
entire functions}\label{nonintegral}

\begin{figure}
	\includegraphics[scale=0.45]{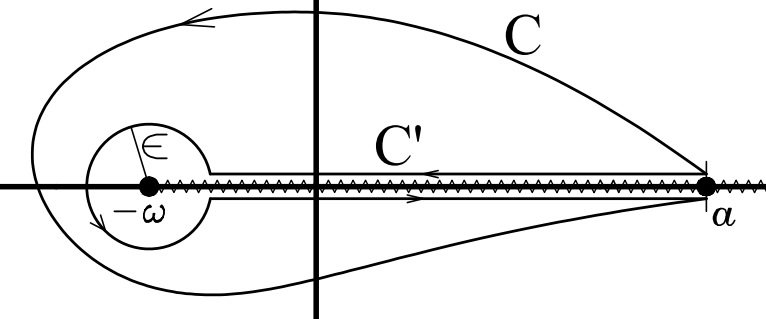}
	\caption{The contour of integration.}
	\label{around_-omega}
\end{figure}

We now perform finite part integration of the generalized Stieltjes transform \eqref{ogiy}  of non-integral order $\lambda=n+\alpha$ 
\begin{equation}\label{blag}
S_{n+\alpha}^a[f]=\int_0^a \frac{f(x)}{(\omega+x)^{n+\alpha}} \mathrm{d}x, \;\; 0<\alpha<1, \; n=1, 2, 3, \dots 
\end{equation}
assuming that the complex extension, $f(z)$, of $f(x)$ is entire. Following on the initial discussion in the Introduction on writing \eqref{blag} as a complex contour integral, 
\begin{equation}
\int_{\mathrm{C}} \frac{f(z)}{(\omega + z)^{n+\alpha}} \mathrm{d}z ,
\end{equation}
we take the branch cut of $(\omega+z)^{-\alpha}$ to be the line $[-\omega,\infty)$.  The contour $\mathrm{C}$ is then deformed into the equivalent contour $\mathrm{C}'$ as depicted in Figure-\ref{around_-omega}. Since $f(z)$  is entire and the deformation of the contour does not pass through any singularity, the integrals along $\mathrm{C}$ and $\mathrm{C}'$ are equal. Hence,
\begin{eqnarray}\label{eq:7}
\int_{\mathrm{C}^\prime}\frac{f(z)}{(\omega+z)^{n+\alpha}}\mathrm{d}z&=&\int_a^{-\omega+\epsilon}\frac{f(x)}{(\omega+x)^{n+\alpha}}\mathrm{d}x + \int_{\epsilon}\frac{f(z)}{(\omega+z)^{n+\alpha}}\mathrm{d}z\nonumber\\\nonumber
&&+\int_{-\omega+\epsilon}^a\frac{f(x)}{(\omega+x)^{n+\alpha}e^{2\pi i(n+\alpha)}}\mathrm{d}x\\\nonumber
&=&\int_a^{-\omega+\epsilon}\frac{f(x)}{(\omega+x)^{n+\alpha}}\mathrm{d}x+\int_{\epsilon}\frac{f(z)}{(\omega+z)^{n+\alpha}}\mathrm{d}z\\\nonumber 
&&+e^{-2\pi i\alpha}\int_{-\omega+\epsilon}^a\frac{f(x)}{(\omega+x)^{n+\alpha}}\mathrm{d}x\nonumber \\
&=&(e^{-2\pi i\alpha}-1)\int_{-\omega+\epsilon}^a\frac{f(x)}{(\omega+x)^{n+\alpha}}\mathrm{d}x+\int_{\epsilon}\frac{f(z)}{(\omega+z)^{n+\alpha}}\mathrm{d}z
\end{eqnarray}

Let us now consider the integral along the small circular contour. Along this contour, we have the parametrization $z=-\omega+\epsilon e^{i\theta}$ for $0<\theta<2\pi$. Since $f(z)$ is entire, we can expand it about any point, in particular, at $z=-\omega$,
	\begin{equation}
    	f(z)=\sum_{j=0}^{n-1}\frac{f^{(j)}(-\omega)}{j!}(z+\omega)^j+R_n(z)
	\end{equation}
where $R_n(z)=O((z+\omega)^n)=O(\epsilon^{n} e^{i n\theta})$. Substituting this expansion back into the integral around the small circular contour, we obtain
\begin{eqnarray}\nonumber
\int_{\epsilon}\frac{f(z)}{(\omega+z)^{n+\alpha}}\mathrm{d}z &=&\int_{0}^{2\pi}\frac{1}{(\epsilon e^{i\theta})^{n+\alpha}}\left[\sum_{j=0}^{n-1}\frac{f^{(j)}(-\omega)}{j!}(\epsilon e^{i\theta})^j+R_n(z)\right]\epsilon e^{i\theta}i\mathrm{d}\theta\\
&=&\sum_{j=0}^{m-1}\frac{i f^{(j)}(-\omega)}{j!\epsilon^{n+\alpha-j-1}}\int_0^{2\pi}e^{-i\theta(n+\alpha-j-1)}\mathrm{d}\theta \nonumber \\
&&\hspace{32mm}+i\int_0^{2\pi}\frac{R_n(z)e^{-i\theta(n+\alpha-1)}}{\epsilon^{n+\alpha-1}}\mathrm{d}\theta
\end{eqnarray}
The integral in the first term above evaluates to
\begin{equation*}
\int_0^{2\pi}e^{-i\theta(n+\alpha-j-1)}\mathrm{d}\theta=\frac{e^{-2\pi i(n+\alpha-j-1)}-1}{-i(n+\alpha-j-1)},
\end{equation*}
while a bound for the second term can be obtained as follows
\begin{eqnarray*}
\left |i\int_0^{2\pi}\frac{R_n(z)e^{-i\theta(n+\alpha-1)}}{\epsilon^{n+\alpha-1}}\mathrm{d}\theta\right|&\le&\epsilon^{1-\alpha}\int_0^{2\pi}\left |\frac{R_n(z)e^{-i\theta(n+\alpha-1)}}{\epsilon^{n}}\right|\mathrm{d}\theta\\
&\le&\epsilon^{1-\alpha}\int_0^{2\pi}\left|\frac{K(\epsilon e^{i\theta})^n e^{-i\theta(n+\alpha-1)}}{\epsilon^n}\right|\mathrm{d}\theta\\
&=&2\pi K\epsilon^{1-\alpha},
\end{eqnarray*}
where $K$ is some positive constant. Thus the integral on the circular loop centered at $z=-\omega$ can be written as 
\begin{equation*}
\int_{\epsilon}\frac{f(z)}{(\omega+z)^{n+\alpha}}\mathrm{d}z=\sum_{j=0}^{n-1}\frac{i\,f^{(j)}(-\omega)(e^{-2\pi i\alpha}-1)}{-i\,j!(n+\alpha-j-1)\epsilon^{n+\alpha-j-1}}+\mathcal{O}\left(\epsilon^{1-\alpha}\right)
\end{equation*}
The integral \eqref{eq:7} along the deformed contour $ \mathrm{C}^\prime$ may now be written as.
\begin{eqnarray*}
\int_{C^\prime}\frac{f(z)}{(\omega+z)^{n+\alpha}}\mathrm{d}z=(e^{-2\pi i\alpha}-1)\left [\int_{-\omega+\epsilon}^{0}\frac{f(x)}{(\omega+x)^{n+\alpha}}\mathrm{d}x+\int_{0}^{a}\frac{f(x)}{(\omega+x)^{n+\alpha}}\mathrm{d}x\right ]\\
-\sum_{j=0}^{m-1}\frac{f^{(j)}(-\omega)(e^{-2\pi i\alpha}-1)}{j!(n+\alpha-j-1)(\epsilon^{n+\alpha-j-1})}+\mathcal{O}\left(\epsilon^{1-\alpha}\right)
\end{eqnarray*}
We rewrite the first term above with the replacement $x\to -x$. Multiplying through by $(e^{-2\pi i\alpha}-1)^{-1}$ and rearranging the terms, we obtain
\begin{eqnarray}
\int_0^a \frac{f(x)}{(\omega + x)^{n+\alpha}} \mathrm{d}x& =& \frac{1}{e^{-2\pi i\alpha}-1}\int_{C\prime}\frac{f(z)}{(\omega+z)^{n+\alpha}}\mathrm{d}z \nonumber \\
&&\hspace{-34mm} - \left[\int_{0}^{\omega-\epsilon}\frac{f(-x)}{(\omega-x)^{n+\alpha}}\mathrm{d}x-\sum_{j=0}^{n-1}\frac{f^{(j)}(-\omega)}{j!(n+\alpha-j-1)(\epsilon^{n+\alpha-j-1})}\right] + \mathcal{O}(\epsilon^{1-\alpha})
\end{eqnarray}

The first term is an integral along the contour $\mathrm{C}'$, which depends on $\epsilon$. However, the contour $\mathrm{C'}$ can be replaced with the contour $\mathrm{C}$ because the two integrals are equal. The first term is then independent of $\epsilon$. Now the left hand side is independent of $\epsilon$ so that the limit of the right hand side exists as $\epsilon\rightarrow 0$. Then we obtain the representation
\begin{eqnarray}
\int_0^a \frac{f(x)}{(\omega + x)^{n+\alpha}} \mathrm{d}x& =& \frac{1}{e^{-2\pi i\alpha}-1}\int_{\mathrm{C}}\frac{f(z)}{(\omega+z)^{n+\alpha}}\mathrm{d}z \nonumber \\
&&\hspace{-28mm} - \lim_{\epsilon\rightarrow 0} \left[\int_{0}^{\omega-\epsilon}\frac{f(-x)}{(\omega-x)^{n+\alpha}}\mathrm{d}x-\sum_{j=0}^{n-1}\frac{f^{(j)}(-\omega)}{j!(n+\alpha-j-1)\epsilon^{n+\alpha-j-1}}\right] 
\end{eqnarray}
From equation \eqref{asol}, we recognize that the limit term is just the finite part of the divergent integral
\begin{equation}
\int_0^{\omega} \frac{f(-x)}{(\omega-x)^{n+\alpha}} \mathrm{d}x
\end{equation}
That is
\begin{equation*}
\bbint{0}{\omega}\frac{f(-x)}{(\omega-x)^{n+\alpha}}\mathrm{d}x=\lim_{\epsilon\rightarrow0}\left[\int_{0}^{\omega-\epsilon}\frac{f(-x)}{(\omega-x)^{n+\alpha}}\mathrm{d}x-\sum_{j=0}^{n-1}\frac{f^{(j)}(-\omega)}{j!(n+\alpha-j-1)(\epsilon^{n+\alpha-j-1})}\right]
\end{equation*}

Finally, a representation of the integral \eqref{blag} in terms of a complex contour integral and a finite part integral is given by,
\begin{equation}\label{new}
\int_{0}^{a}\frac{f(x)}{(\omega+x)^{n+\alpha}}\mathrm{d}x=\frac{1}{e^{-2\pi i\alpha}-1}\int_{\mathrm{C}}\frac{f(z)}{(\omega+z)^{n+\alpha}}\mathrm{d}z-\bbint{0}{\omega}\frac{f(-x)}{(\omega-x)^{n+\alpha}}\mathrm{d}x .
\end{equation}
The second term in the right hand side of equation \eqref{new} is represented as a complex contour integral in equation \eqref{oit}.
\begin{align}\nonumber
    \bbint{0}{\omega}\frac{f(-x)}{(\omega-x)^{n+\alpha}}\mathrm{d}x &= \bbint{0}{\omega}\frac{f(x-\omega)}{x^{n+\alpha}}\mathrm{d}x\\
    &= \frac{1}{e^{-2\pi\alpha i}-1}\int_{\mathrm{C}}\frac{f(z-\omega)}{z^{n+\alpha}}\mathrm{d}z
\end{align}
Hence, the right hand side of equation (\ref{new}) is a complex contour integral representation of the given convergent integral in the left hand side.
As for the first term, we are now ready to implement a term by term integration with the use of the following binomial expansion
\begin{eqnarray}
\frac{1}{\left(\omega+z\right)^{n+\alpha}}=\frac{1}{z^{n+\alpha}}\sum_{j=0}^{\infty}{{-n-\alpha}\choose{j}}\left(\frac{\omega}{z}\right)^{j}
\end{eqnarray}
This expansion converges provided $\omega<|z|$. We can substitute this back in the integral as long as we choose the contour such that $\omega<|z|$ for all $z$ in the contour $\mathrm{C}$. Under this condition, the infinite series converges uniformly along the contour of integration, allowing us to perform a term by term integration, yielding the finite part integral representation of the given Stieltjes transform,
\begin{eqnarray*}
\int_{0}^{a}\frac{f(x)}{(\omega+x)^{n+\alpha}}\mathrm{d}x=\sum_{j=0}^{\infty}{{-n-\alpha}\choose{j}}\omega^{j}\bbint{0}{a}\frac{f(x)}{x^{n+\alpha+j}}\mathrm{d}x-\bbint{0}{\omega}\frac{f(-x)}{(\omega-x)^{n+\alpha}}\mathrm{d}x
\end{eqnarray*}
The explicit condition through which the infinite series in the right hand side of the equation above converges absolutely is obtained by deforming the contour of integration $\mathrm{C}$ in Figure-\ref{around_-omega} into a circular path with radius $a$. We then proceed in a similar fashion as in\cite[eq.63 to 64]{tica} to establish that the absolute convergence of the series is ensured for $\omega<a$. Thus we have proved the following result.

\begin{theorem}\label{branch}
Let $f(x)$ be locally integrable in the interval $[0,a]$ with an entire complex extension $f(z)$. Then
	\begin{equation}\label{new_rezult}
	\int_0^a\frac{f(x)}{(\omega+x)^{n+\alpha}}\mathrm{d}x=\sum_{j=0}^{\infty}{{-n-\alpha}\choose {j}}\, {\omega}^j \, \bbint{0}{a}\frac{f(x)}{x^{n+\alpha+j}}\mathrm{d}x + \Delta_{\mathrm{sc}}^{(n+\alpha)}(\omega),
	\end{equation}
where $n=1,2,3...$ and $0<\alpha<1$, provided $\omega<a$, in which the singular contribution is the finite part integral
\begin{eqnarray}\label{singular_terrm}
\Delta_{\mathrm{sc}}^{(n+\alpha)}(\omega)&=&- \bbint{0}{\omega} \frac{f(-x)}{(\omega-x)^{n+\alpha}}\mathrm{d}x = -\bbint{0}{\omega}\frac{f(x-\omega)}{x^{n+\alpha}}\mathrm{d}x \nonumber \\
&=& - \sum_{j=0}^{\infty}\frac{f^{\left(j\right)}\left(-\omega\right) \,\omega^{j-n-\alpha+1}}{j!\,\left(j+1-n-\alpha\right)} .
\end{eqnarray}
\end{theorem}

\section{Behavior for small parameters}\label{param}
The dominant contribution to the value of $S_{n+\alpha}^{a}\left[f\right]$ is determined by the order of the zero of $f$ at the origin relative to the order of the transform, $n+\alpha$. For a function with a zero of order $m=n-s$, for $s=1,2,\dots, n$, we write $f\left(z\right) = \sum_{k=0}^{\infty}d_{k}\,z^{k+n-s}$. The singular term \eqref{singular_terrm} provides the dominant contribution of the order $\mathcal{O}\left(\omega^{-s-\alpha+1}\right)$ given by
\begin{eqnarray}\nonumber
\Delta_{\mathrm{sc}}^{(n+\alpha)}\left(\omega\right) &=& \sum_{j=0}^{n-s}\sum_{k=0}^{\infty}\frac{d_k\,\left(k+n-s\right)!\,\left(-1\right)^{k+n-s-j}\,\omega^{k-s-\alpha+1}}{j!\,\left(k+n-s-j\right)!\,\left(n+\alpha-j-1\right)}\\
&-&\sum_{j=n-s+1}^{\infty}\sum_{k=j-n+s}^{\infty}\frac{d_k\,\left(k+n-s\right)!\,\left(-1\right)^{k+n-s-j}\,\omega^{k-s-\alpha+1}}{j!\,\left(k+n-s-j\right)!\,\left(j-n-\alpha+1\right)}
\end{eqnarray}
So that up to leading order,
\begin{equation}\label{ort}
\int_{0}^{a}\frac{f\left(x\right)}{\left(\omega+x\right)^{n+\alpha}}\mathrm{d}x \sim \frac{d_0\,\left(n-s\right)!}{\omega^{s+\alpha-1}}\sum_{j=0}^{n-s}\frac{\left(-1\right)^{n-s-j}}{j!\,\left(n-s-j\right)!\,\left(n+\alpha-j-1\right)},\,\,\omega\to 0 .
\end{equation}

On the other hand, for a function $f$ with a zero at the origin of order $m=n+r$ for $r=0,1,2,\dots$, we write $f\left(z\right) = \sum_{k=0}^{\infty}d_k z^{k+n+r}$. We find that in this case, the naive term provides the dominant contribution to the value of $S_{n+\alpha}^{a}\left[f\right]$ while the singular term \eqref{singular_terrm} only gives a leading order correction given by,
\begin{eqnarray}\nonumber
\Delta_{\mathrm{sc}}^{(n+\alpha)}\left(\omega\right) &=& \sum_{j=0}^{n+r}\sum_{k=0}^{\infty}\frac{d_k\,\left(k+n+r\right)!\,\left(-1\right)^{k+n+r-j}\,\omega^{k+r-\alpha+1}}{j!\,\left(k+n+r-j\right)!\,\left(n+\alpha-j-1\right)}\\
&-&\sum_{j=n+r+1}^{\infty}\sum_{k=j-n-r}^{\infty}\frac{d_k\,\left(k+n+r\right)!\,\left(-1\right)^{k+n+r-j}\,\omega^{k+r-\alpha+1}}{j!\,\left(k+n+r-j\right)!\,\left(j-n-\alpha+1\right)}
\end{eqnarray}
In this case, the leading term corresponding to $j=0$ in the naive term provides the dominant contribution 
\begin{equation}
\int_{0}^{a}\frac{f\left(x\right)}{\left(\omega+x\right)^{n+\alpha}}\mathrm{d}x \sim \bbint{0}{a}\frac{f\left(x\right)}{x^{n+\alpha}}\mathrm{d}x,\,\,\omega\to 0
\end{equation}
Since the order of the zero of $f$ at the origin is greater than or equal to $n$, the leading finite part integral above is a convergent integral. In fact, the same can be said for the finite part integrals corresponding to $0\leq j\leq r$ in the naive term of \eqref{new_rezult}. The singular contributions begin to provide significant correction at $j=r+1$.

\section{Examples}\label{ung}
\subsection{Example}\label{pina}
We apply Theorem-\ref{branch} to the following specialized values of the Gauss hypergeometric function
\begin{align}\label{integraal}
\pFq{2}{1}{n+\alpha,r}{s}{-z} = \frac{\left(s-1\right)!}{\left(r-1\right)!\,\left(s-r-1\right)!\,z^{n+\alpha}}\,\int_{0}^{1}\frac{x^{r-1}\,\left(1-x\right)^{s-r-1}}{\left(z^{-1}+x\right)^{n+\alpha}}\mathrm{d}x,
\end{align}
for positive integers $s,r,n$ and $0<\alpha<1$ with $s\geq r+1$ and $z=\zeta>1$, to reproduce a known representation of a class of $_{p}F_q$ \cite{simplepole2F1}.

From Theorem-\ref{branch}, the integral is evaluated as 
\begin{align}\label{gauss_app}
\int_{0}^{1}\frac{x^{r-1}\left(1-x\right)^{s-r-1}}{\left(\zeta^{-1}+x\right)^{n+\alpha}}\mathrm{d}x &= \sum_{j=0}^{\infty}{{-n-\alpha}\choose j}\frac{1}{\zeta^{j}}\,\bbint{0}{1}\frac{x^{r-1}\left(1-x\right)^{s-r-1}}{x^{n+\alpha+j}}\mathrm{d}x\\\nonumber
&-\bbint{0}{\zeta^{-1}}\frac{\left(-x\right)^{r-1}\,\left(1+x\right)^{s-r-1}}{\left(\zeta^{-1}-x\right)^{n+\alpha}}\mathrm{d}x
\end{align}
The finite part in the first term of the right hand side of \eqref{gauss_app} is a specific case of \cite[eq.53]{tica},
\begin{align}\label{first_FP}
\bbint{0}{1}\frac{x^{r-1}\left(1-x\right)^{s-r-1}}{x^{n+\alpha+j}}\,\mathrm{d}x = \sum_{k=0}^{s-r-1}\frac{\left(s-r-1\right)!\,\left(-1\right)^{k}}{k!\,\left(s-r-1-k\right)!\,\left(k-n-\alpha-j+r\right)}
\end{align}
while the finite part in the second term is obtained using the steps leading to equation \eqref{nit} with the identification $g\left(x\right)=\left(-x\right)^{r-1}\left(1+x\right)^{s-r-1}$. Making use of the following Taylor expansion about $x = \zeta^-1$
\begin{align}\label{koi}
\left(-x\right)^{r-1}\left(1+x\right)^{s-r-1} = \sum_{k=0}^{s-2}\frac{M_{k}\left(\zeta^{-1}\right)}{k!}\left(x-\zeta^{-1}\right)^{k},
\end{align}
where 
\begin{equation}
M_{k}\left(\zeta^{-1}\right) 
=\left(-1\right)^{r-1}\,\left(s-r-1\right)!\,\sum_{j=k}^{s-2}\frac{j!\,(\zeta^{-1})^{j-k}}{(j-r+1)! (s-j-2)! (j-k)!},
\end{equation}
the second finite part is computed as 
\begin{equation}\label{second_FP}\nonumber
\bbint{0}{\zeta^{-1}}\frac{\left(-x\right)^{r-1}\,\left(1+x\right)^{s-r-1}}{\left(\zeta^{-1}-x\right)^{n+\alpha}}\mathrm{d}x = \sum_{k=0}^{s-2}\frac{M_{k}\left(\zeta^{-1}\right)\,\left(-1\right)^{k}}{k!\,\left(k+1-n-\alpha\right)}\left(\zeta^{-1}\right)^{k+1-n-\alpha}
\end{equation}
Interchanging the order of summation according to
\begin{equation}
	\sum_{k=0}^{p}\sum_{j=k}^{p}a_{k,j} = \sum_{j=0}^{p}\sum_{k=0}^{j}a_{k,j}
\end{equation}
and simplifying, we obtain
\begin{equation}\label{piy}
	\bbint{0}{\zeta^{-1}}\frac{\left(-x\right)^{r-1}\,\left(1+x\right)^{s-r-1}}{\left(\zeta^{-1}-x\right)^{n+\alpha}}\mathrm{d}x = \frac{(s-r-1)!}{\zeta^{1-n-\alpha}}\sum_{j=r-1}^{s-2}\frac{(-1)^{r-1}\,m_j\,j!}{(j-r+1)!\,(s-j-2)!\,\zeta^{j}}
\end{equation}
where the inner sum evaluates to
\begin{equation}
	m_j = \sum_{k=0}^{j}\frac{(-1)^{k}}{k! (k+1-n-\alpha)\,(j-k)!} = \frac{\Gamma(1-n-\alpha)}{\Gamma(2+j-n-\alpha)}.
\end{equation}

The index of summation in equation \eqref{piy} can be shifted $j\to j-r+1$ and then substituted, along with the result in equation
 \eqref{first_FP}, into \eqref{gauss_app}. The resulting expression can then be substituted for the integral in \eqref{integraal}, to obtain the following expansion for the Gauss hypergeometric function
\begin{align}\label{wito}
\pFq{2}{1}{n+\alpha,r}{s}{-\zeta} = \frac{\left(s-1\right)!\,\Gamma\left(1-n-\alpha\right)}{\left(r-1\right)!\,\zeta^{n+\alpha}}\sum_{j=0}^{\infty}\frac{b_{j}}{\Gamma\left(1-n-\alpha-j\right)}\frac{\zeta^{-j}}{j!}\\\nonumber
+\frac{(-1)^{r}(s-1)!\,\Gamma(1-n-\alpha)}{(r-1)!\,\zeta^r}\sum_{j=0}^{s-r-1}\frac{(j+r-1)!}{j!(s-j-r-1)!\,\Gamma(j+r-n-\alpha+1)\,\zeta^{j}}
\end{align}
where 
\begin{equation}
b_{j} = \sum_{k=0}^{s-r-1}\frac{\left(-1\right)^{k}}{k!\,\left(s-r-1-k\right)!\,\left(k-n-\alpha-j+r\right)} = \frac{\Gamma(r-n-\alpha-j)}{\Gamma(s-n-\alpha-j)}
\end{equation}
for all $0<\alpha<1$, $\zeta>1$, $n=1, 2, 3, \dots$, $r=1, 2, 3, \dots$ and $s=(r+1), (r+2), (r+3), \dots$. The representation \eqref{wito} gives the following relation for the Gauss hypergeometric function
\begin{align}\nonumber
\pFq{2}{1}{n+\alpha,r}{s}{-\zeta} =& \frac{(s-1)!\,\Gamma(r-n-\alpha)}{(r-1)!\,\Gamma(s-n-\alpha)\,\zeta^{n+\alpha}}\pFq{2}{1}{n+\alpha,n+\alpha-s+1}{n+\alpha-r+1}{-\frac{1}{\zeta}}\\
&+ \frac{(-1)^{r}(s-1)!\,\Gamma(1-n-\alpha)}{(s-r-1)!\,\Gamma(r+1-n-\alpha)\,\zeta^{r}}\pFq{2}{1}{r,r-s+1}{r-n-\alpha+1}{-\frac{1}{\zeta}}\label{momo}
\end{align}
The right hand side of equation \eqref{momo} serves as an analytic extension of the infinite series representation of the left hand side for $\zeta>1$. These results are identical to the tabulated result for the hypergeometric function  given in \cite{simplepole2F1} for the case of a simple pole.

Moreover, When $n=3$, $r=2$, $s=3$, and $\alpha=\frac{1}{2}$ the expansion \eqref{wito} reduces to the following special case given in \cite[p482,\#263]
{prudnikov}
\begin{equation}
\pFq{2}{1}{\frac{7}{2},2}{3}{-\zeta} = \frac{4}{15\zeta^2}\left[2-\left(2+5\zeta\right)\left(1+\zeta\right)^{-\frac{5}{2}}\right].
\end{equation}
We also obtain the following asymptotic expansion
\begin{eqnarray}\nonumber
\pFq{2}{1}{n+\alpha,r}{s}{-\zeta} &=&\frac{(s-1)!\,\Gamma(r-n-\alpha)}{(r-1)!\,\Gamma(s-n-\alpha)\zeta^{n+\alpha}}\left(1+\mathcal{O}\left(\frac{1}{\zeta}\right)\right)\\
&&+ \frac{(-1)^{r}\,(s-1)!\,\Gamma(1-n-\alpha)}{(s-r-1)!\,\Gamma(r-n-\alpha+1)\,\zeta^{r}}\left(1+\mathcal{O}\left(\frac{1}{\zeta}\right)\right)
\end{eqnarray}
as $\zeta\to\infty$. This asymptotic behavior is also tabulated in \cite{simplepole2F1}.
\subsection{Example}
We obtain a series representation of the Kummer function of the second kind for the specialized values $z=\omega>0$, $a=n=1, 2, \dots$ and $b=1-\alpha$ for $0<\alpha<1$. We cast the integral representation to assume a form of a Stieltjes transform by changing the variable to $x=\omega t$. This leads to the specialized integral representation
\begin{equation}\label{owt}
U(n,1-\alpha,\omega) = \frac{\omega^{\alpha}}{(n-1)!}\int_0^{\infty}\frac{e^{-x}x^{n-1}}{(\omega+x)^{n+\alpha}}\, \mathrm{d}x
\end{equation}
The integral is now in a form amenable to application of Theorem-\ref{branch}. We identify $f(x)=e^{-x} x^{n-1}$. From Theorem-\ref{branch}, the integral assumes the expansion
\begin{equation}\label{theorem}
\int_{0}^{\infty}\frac{e^{-x}x^{n-1}}{\left(\omega+x\right)^{n+\alpha}}\,\mathrm{d}x = \sum_{j=0}^{\infty}{{-n-\alpha}\choose{j}}\omega^{j}\bbint{0}{\infty}\frac{e^{-x}}{x^{j+\alpha+1}}\mathrm{d}x-\bbint{0}{\omega}\frac{e^{x}\left(-x\right)^{n-1}}{\left(\omega-x\right)^{n+\alpha}}\mathrm{d}x .
\end{equation}

The finite part integral in the first term of the right hand side of equation \eqref{theorem} is a specialized value of the finite part integral given in \cite[eq.48]{tica}. The value is given by
\begin{equation}
\bbint{0}{\infty}\frac{e^{-x}}{x^{j+\alpha+1}}\mathrm{d}x = \frac{\left(-1\right)^{j+1}\,\pi}{\sin\left(\pi\,\alpha\right)\,\Gamma\left(j+\alpha+1\right)} .
\end{equation}

We compute the finite part integral in the second term using the procedure leading to  \eqref{nit}. With the identification that $g(x)=(-x)^{n-1} e^x$, the required expansion is given by
\begin{equation}\label{series}
e^{x}\left(-x\right)^{n-1}=\sum_{k=0}^{\infty}\frac{M_{k}\left(\omega\right)}{k!}\left(x-\omega\right)^k
\end{equation}
where
\begin{equation}\label{coefficient}
M_{k}\left(\omega\right) = e^{\omega}\left(-1\right)^{n-1}\sum_{l=0}^{k}\frac{k!}{l!\,\left(k-l\right)!}\frac{\left(n-1\right)!\,\,\omega^{n-1-l}}{\left(n-l-1\right)!}.
\end{equation}
 Hence, the finite part is given by
\begin{eqnarray}\nonumber
\bbint{0}{\omega}\frac{e^{x}\left(-x\right)^{n-1}}{\left(\omega-x\right)^{n+\alpha}}\,\mathrm{d}x&=&\sum_{k=0}^{\infty}\frac{\left(-1\right)^{k}M_{k}\left(\omega\right)}{k!\left(k-\alpha-n+1\right)}\omega^{k+1-n-\alpha}\nonumber\\
&=&e^{\omega}\left(-1\right)^{n-1}\omega^{-\alpha}\left(n-1\right)!\nonumber \\
&& \hspace{8mm} \times \sum_{k=0}^{\infty}\sum_{l=0}^{k}\frac{\left(-1\right)^{k}\omega^{k-l}}{\left(k-\alpha-n+1\right)l!\left(k-l\right)!\left(n-l-1\right)!}
\end{eqnarray}
where we arrived at the second line by substituting the coefficient \eqref{coefficient} in the first line. We simplify the double sum by applying the identity
\begin{equation}
\sum_{k=0}^{\infty}\sum_{l=0}^{k}a_{k,l}=\sum_{l=0}^{\infty}\sum_{k=l}^{\infty}a_{k,l}
\end{equation}
followed by the substitution $k-l=r$ and by an interchange of the order of summation. The result is
\begin{eqnarray}\nonumber\label{kiso}
\bbint{0}{\omega}\frac{e^{x}\left(-x\right)^{n-1}}{\left(\omega-x\right)^{n+\alpha}}\,\mathrm{d}x
&=& \frac{\left(-1\right)^{n-1}e^{\omega}\left(n-1\right)!}{\omega^{\alpha}}\nonumber \\
&&\times\sum_{r=0}^{\infty}\sum_{l=0}^{\infty}\frac{\left(-1\right)^{r+l}\,\omega^r}{\left(r+l-\alpha-n+1\right)\,l!\,r!\,\left(n-l-1\right)!}
\end{eqnarray}
Finally the inner summation can be evaluated as
\begin{equation}
\sum_{l=0}^{\infty}\frac{\left(-1\right)^{l}}{\left(r+l-\alpha-n+1\right)\,l!\,\left(n-l-1\right)!}=\frac{\Gamma\left(1-\alpha-n+r\right)}{\Gamma\left(1-\alpha+r\right)}.
\end{equation}
Collecting the results above, an exact expansion of Kummer function of the second kind for the specified parameter values is given by
\begin{eqnarray}\label{opi}
U\left(n,1-\alpha,\omega\right)
&=&-\frac{\pi\,\omega^{\alpha}\,\Gamma\left(1-n-\alpha\right)}{\sin\left(\pi\,\alpha\right)\,\left(n-1\right)!}\,\sum_{j=0}^{\infty}\frac{\left(-1\right)^{j}}{\Gamma\left(1-n-\alpha-j\right)\,\Gamma\left(j+\alpha+1\right)}\,\frac{\omega^{j}}{j!}\nonumber\\
&&+\left(-1\right)^{n}e^{\omega}\sum_{r=0}^{\infty}\frac{\left(-1\right)^{r}\Gamma\left(1-\alpha-n+r\right)}{\Gamma\left(1-\alpha+r\right)}\,\frac{\omega^{r}}{r!}
\end{eqnarray}
for all $0<\alpha<1$, $n=1, 2, 3, \dots$ and $\omega>0$. The expansion above can then be written in the following form
\begin{align}\nonumber\label{onu}
U\left(n,1-\alpha,\omega\right) 
=& -\frac{\pi\omega^{\alpha}}{\sin(\pi\alpha)\,(n-1)!\,\Gamma(1+\alpha)}\pFq{1}{1}{n+\alpha}{1+\alpha}{\omega}\\
& + \frac{(-1)^{n}e^{\omega}\Gamma(1-n-\alpha)}{\Gamma(1-\alpha)}\pFq{1}{1}{1-n-\alpha}{1-\alpha}{-\omega}.
\end{align}When $n=2$, $\alpha=1/2$, the expansion \eqref{opi} reduces to the following tabulated special case given in \cite{kummer2}, 
\begin{equation}
U(2,\frac{1}{2},\omega) = -\frac{2}{3}\left[\sqrt{\pi\omega}e^{\omega}\left(2\omega+3\right)\mathrm{erfc}(\omega)-2(\omega+1)\right]
\end{equation}
where $\mathrm{erfc(\omega)}$ is the complementary error function. 
It is worth noting that the left hand side of the representations \eqref{opi} and \eqref{onu}, is not amenable to symbolic evaluation routines in computer algebra systems such as Mathematica or Maple for a general positive integer $n$ given a specific $\alpha$ and $\omega$. In contrast, one could make use of the  right hand side of these expansions, to obtain representations of the Kummer function involving the unspecified parameter $n$. 

 Moreover, the following leading order behavior for $\omega\ll 1$ can then be directly extracted from the exact representation \eqref{opi},
\begin{eqnarray}\nonumber
U\left(n,1-\alpha,\omega\right) &=& -\frac{\omega^{\alpha}\,\Gamma(\alpha)}{(n-1)!}\left(1+\mathcal{O}(\omega)\right)+\frac{(-1)^{n}\,e^{\omega}\,\Gamma(1-\alpha-n)}{\Gamma(1-\alpha)}\left(1+\mathcal{O}\left(\omega\right)\right)\\\label{keri}
&=& -\frac{\omega^{\alpha}\,\Gamma(\alpha)}{(n-1)!}\left(1+\mathcal{O}(\omega)\right)+\frac{e^{\omega}\Gamma(\alpha)}{\Gamma(\alpha+n)}\left(1+\mathcal{O}(\omega)\right)
\end{eqnarray}
The dominant contribution in the expansion above comes from the singular term \eqref{kiso} as expected from the order of the zero of $f(x)= x^{n-1}e^{-x}$ at the origin which is less than the order $n+\alpha$ of the Stieltjes transform. The leading  term in \eqref{keri} is of the order $\mathcal{O}\left(\omega^0\right)$ since the integral factor in the representation \eqref{owt} has a leading behavior of $\mathcal{O}\left(\omega^{-\alpha}\right)$ according to equation \eqref{ort}.

We may also compare our result with the following limiting behavior given in \cite[p508, eq.13.5.10]{abramowitz}
\begin{equation}
U(a,b,z) = \frac{\Gamma(1-b)}{\Gamma(1+a-b)} + \mathcal{O}(|z|^{1-\mathrm{Re}(b)}),\,\,\,0<\mathrm{Re}(b)<1.
\end{equation}
This corresponds to the dominant singular term in equation \eqref{keri} with the sub-dominant terms omitted altogether and the exponential factor replaced by unity.

\section{A Further Generalization of Stieltjes Transform}\label{appli}
In this section, we consider a particular generalization of the generalized Stieltjes transform \eqref{ogiy} given by 
\begin{equation}\label{yit}
F(\omega) = \int_{0}^{\infty}\frac{f(x)}{\sqrt{\omega^{2}+x^{2}}}\mathrm{d}x,\qquad \omega>0.
\end{equation}
Finite part integration of \eqref{yit} yields an exact representation from which the asymptotic behavior for $\omega\ll 1$ can be directly extracted. Obtaining the asymptotic behavior of the integral in equation \eqref{yit} in this regime is relevant in the calculation of the effective index of refraction of a shallow one dimensional quantum well \cite{galaponprl,alvin}. The finite part integration of \eqref{yit} can be readily extended in the evaluation of transforms of the form
\begin{equation}
F(\omega) = \int_{0}^{\infty}\frac{f(x)}{(\omega^{m}+x^{m})^{\gamma}}\mathrm{d}x,\qquad \omega>0, \;\; m=1, 2, \dots, \;\; \gamma>0 .
\end{equation}
The relevant expansion is given by the following result.
\begin{theorem}\label{orat}
Let $f(x)$ be a function of the real variable $x\in[0,a]$. If $f(x)$ posses an entire complex extension, $f(z)$, then 
\begin{align}\label{pinak}
	\int_{0}^{a}\frac{f(x)}{\sqrt{\omega^{2}+x^{2}}}\mathrm{d}x = \sum_{k=0}^{\infty}{{-\frac{1}{2}\choose{k}}}\omega^{2k}\,\bbint{0}{a}\frac{f(x)}{x^{2k+1}}\mathrm{d}x + \Delta_{sc},
\end{align}  
where the correction term is the singular contributions given by
\begin{align}\nonumber\label{kor}
\Delta_{sc} = -\frac{1}{2\sqrt{\pi}}\sum_{j=0}^{\infty}\frac{(-1)^{j}\,f^{(2j)}(0)\,\Gamma(j+\frac{1}{2})\,\omega^{2j}}{(2j)!\,j!}\left(H_{j-\frac{1}{2}}-H_j+2\ln\omega\right)\hspace{0.8cm} \\
-\frac{\sqrt{\pi}}{2}\sum_{j=0}^{\infty}\frac{(-1)^{j}\,j!\,f^{(2j+1)}(0)\,\omega^{2j+1}}{(2j+1)!\,\Gamma(j+\frac{3}{2})}\hspace{3cm}
\end{align}
in which $H_s=\psi(s+1)+\gamma$ is a harmonic number.
\end{theorem}
\begin{figure}
	\includegraphics[scale=0.5]{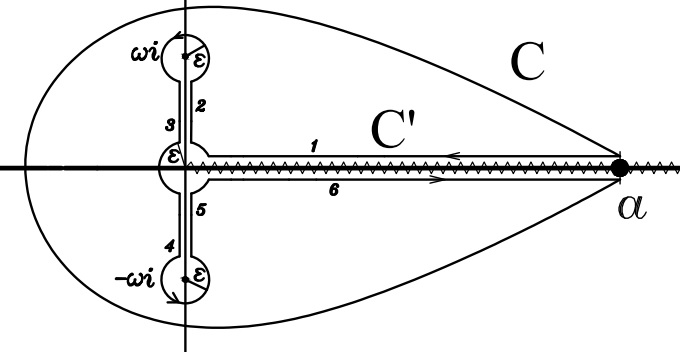}
	\caption{The contour of integration.}
	\label{latest}
\end{figure}
\begin{proof}
Consider the integral evaluated at the contour shown in Figure-\ref{latest}. The contour $\mathrm{C}$ is deformed into the equivalent contour $\mathrm{C'}$.

\begin{align}
\int_{\mathrm{C}}\frac{f\left(z\right)\log z}{\left(\omega^{2}+z^{2}\right)^{\frac{1}{2}}}\mathrm{d}z=\int_{\mathrm{C}'}\frac{f\left(z\right)\log z}{\left(\omega^{2}+z^{2}\right)^{\frac{1}{2}}}\mathrm{d}z
\end{align}
Along $l_{6}$, $z=x\,e^{2\pi\,i}$ 
\begin{align}
\int_{l_{6}}\frac{f\left(z\right)\log z}{\left(\omega^{2}+z^{2}\right)^{\frac{1}{2}}}\mathrm{d}z=\int_{\epsilon}^{a}\frac{f\left(x\right)\ln x}{\sqrt[]{\omega^{2}+x^{2}}}\mathrm{d}x
+\left(2\pi i\right)\int_{\epsilon}^{a}\frac{f\left(x\right)}{\sqrt[]{\omega^{2}+x^{2}}}\mathrm{d}x
\end{align}
along $l_{1}$, $z=x$
\begin{align}
\int_{l_{1}}\frac{f\left(z\right)\log z}{\left(\omega^{2}+z^{2}\right)^{\frac{1}{2}}}\mathrm{d}z=\int_{a}^{\epsilon}\frac{f\left(x\right)\ln x}{\sqrt[]{\omega^{2}+x^{2}}}\mathrm{d}x
\end{align}
along $l_{2}$ and $l_{5}$, $z=i\,y$. Hence along $l_{2}$, 
\begin{align}
\int_{l_{2}}\frac{f\left(z\right)\log z}{\left(\omega^{2}+z^{2}\right)^{\frac{1}{2}}}\mathrm{d}z=\int_{\epsilon}^{\omega-\epsilon}\frac{f\left(iy\right)\log\left(iy\right)\,i}{\sqrt[]{\omega^{2}-y^{2}}}\mathrm{d}y
\end{align}
and $l_{5}$ 
\begin{align}
\int_{l_{5}}\frac{f\left(z\right)\log z}{\left(\omega^{2}+z^{2}\right)^{\frac{1}{2}}}\mathrm{d}z=\int_{-\omega+\epsilon}^{-\epsilon}\frac{f\left(iy\right)\log\left(iy\right)\,i}{\sqrt[]{\omega^{2}-y^{2}}}\mathrm{d}y
\end{align}
along $l_{3}$ and $l_{4}$, $z=i\,y\,e^{2\pi i}$. Hence the integral along $l_{3}$
\begin{align}
\int_{l_{3}}\frac{f\left(z\right)\log z}{\left(\omega^{2}+z^{2}\right)^{\frac{1}{2}}}\mathrm{d}z=\int_{\omega-\epsilon}^{\epsilon}\frac{f\left(iy\right)\log\left(iy\right)\,i}{\sqrt[]{\omega^{2}-y^{2}}\,e^{\pi i}}\mathrm{d}y
\end{align}
and along $l_{4}$
\begin{align}
\int_{l_{4}}\frac{f\left(z\right)\log z}{\left(\omega^{2}+z^{2}\right)^{\frac{1}{2}}}\mathrm{d}z=\int_{-\epsilon}^{-\omega+\epsilon}\frac{f\left(iy\right)\log\left(iy\right)\,i}{\sqrt[]{\omega^{2}-y^{2}}\,e^{\pi i}}\mathrm{d}y
\end{align}
along the upper circular loop, $z=\epsilon\,e^{i\theta}+i\,\omega$ for $-\frac{\pi}{2}<\theta<\frac{3\pi}{2}$. 

We also make use of the Taylor expansion of $f(z)$ about the point $z=i\,\omega$ 
\begin{align}
f(z)=\sum_{k=0}^{\infty}\frac{f^{k}(i\,\omega)}{k!}\left(\epsilon\,e^{i\,\theta}\right)^{k}
\end{align}
so that
\begin{align}
\int_{\epsilon_{1}}\frac{f\left(z\right)\log z}{\left(\omega^{2}+z^{2}\right)^{\frac{1}{2}}}\mathrm{d}z=\int_{-\frac{\pi}{2}}^{\frac{3\pi}{2}}\frac{\log\left(\epsilon\,e^{i\,\theta}+i\,\omega\right)\epsilon\,e^{i\theta}\,i}{\left(\epsilon\,e^{i\,\theta}\right)^{\frac{1}{2}}\left(\epsilon\,e^{i\theta}+2i\,\omega\right)^{\frac{1}{2}}}\sum_{k=0}^{\infty}\frac{f^{k}\left(i\,\omega\right)}{k!}\left(\epsilon\,e^{i\,\theta}\right)^{k}\,\mathrm{d}\theta
\end{align}
Along the lower circular loop, $z=\epsilon\,e^{i\,\theta}-i\,\omega$. We also make use of the Taylor expansion of $f(z)$ about the point $z=-i\,\omega$
\begin{align}
f(z)=\sum_{k=0}^{\infty}\frac{f^{k}(-i\,\omega)}{k!}\left(\epsilon\,e^{i\,\theta}\right)^{k}
\end{align}
Hence
\begin{align}
\int_{\epsilon_{2}}\frac{f\left(z\right)\log z}{\left(\omega^{2}+z^{2}\right)^{\frac{1}{2}}}\mathrm{d}z=\int_{-\frac{3\pi}{2}}^{\frac{\pi}{2}}\frac{\log\left(\epsilon\,e^{i\,\theta}-i\,\omega\right)\epsilon\,e^{i\theta}\,i}{\left(\epsilon\,e^{i\,\theta}\right)^{\frac{1}{2}}\left(\epsilon\,e^{i\theta}-2i\,\omega\right)^{\frac{1}{2}}}\sum_{k=0}^{\infty}\frac{f^{k}\left(-i\,\omega\right)}{k!}\left(\epsilon\,e^{i\,\theta}\right)^{k}\,\mathrm{d}\theta
\end{align}
The contour integrals along the upper and lower loops are of $\mathcal{O}\left(\epsilon^{\frac{1}{2}}\right)$ for small values of $\epsilon$. Hence as $\epsilon\to 0$, these integrals ultimately vanish.

Along the loop centered about the origin, $z=\epsilon\,e^{i\,\theta}$ for $0\le\theta\le2\pi$. We also expand $f(x)$ as a Taylor series about the origin. Hence the integral is given by
\begin{align}
\int_{\epsilon_{3}}\frac{f\left(z\right)\log z}{\left(\omega^{2}+z^{2}\right)^{\frac{1}{2}}}\mathrm{d}z=\int_{0}^{2\pi}\frac{\log\left(\epsilon\,e^{i\theta}\right)\epsilon\,e^{i\theta}\,i}{\sqrt[]{\omega^{2}+\left(\epsilon\,e^{i\theta}\right)^{2}}}\sum_{k=0}^{\infty}\frac{f^{k}\left(0\right)}{k!}\left(\epsilon\,e^{i\theta}\right)^{k}\,\mathrm{d}\theta
\end{align}
the integral above vanishes in the limit as $\epsilon\to 0$ since $\epsilon\to 0$ much faster than $\ln\epsilon\to-\infty$ in the numerator.
Hence after substituting the non-vanishing contributions and simplifying, one obtains 
\begin{align}\nonumber\label{eq:rhs}
\int_{\mathrm{C}'}\frac{f\left(z\right)\log z}{\left(\omega^{2}+z^{2}\right)^{\frac{1}{2}}}\mathrm{d}z&=\left(2\pi i\right)\int_{\epsilon}^{a}\frac{f\left(x\right)}{\sqrt[]{\omega^{2}+x^{2}}}\mathrm{d}x+2\,i\int_{\epsilon}^{\omega-\epsilon}\frac{\ln y\,(f\left(i\,y\right)+f\left(-i\,y\right))}{\sqrt[]{\omega^{2}-y^{2}}}\mathrm{d}y\\
&-\pi\,\int_{\epsilon}^{\omega-\epsilon}\frac{f(i\,y)}{\sqrt[]{\omega^{2}-y^{2}}}\mathrm{d}y-3\pi\,\int_{\epsilon}^{\omega-\epsilon}\frac{f(-i\,y)}{\sqrt[]{\omega^{2}-y^{2}}}\mathrm{d}y
\end{align}
the LHS of \eqref{eq:rhs} can be written as 
\begin{align}\label{eq:lhs}
\int_{\mathrm{C}'}\frac{f\left(z\right)\log z}{\left(\omega^{2}+z^{2}\right)^{\frac{1}{2}}}\mathrm{d}z=\int_{\mathrm{C}'}\frac{f\left(z\right)(\log z-\pi\,i)}{\left(\omega^{2}+z^{2}\right)^{\frac{1}{2}}}\mathrm{d}z+\pi\,i\int_{\mathrm{C}'}\frac{f(z)}{\left(\omega^{2}+z^{2}\right)^{\frac{1}{2}}}\mathrm{d}z
\end{align}
the first term of \eqref{eq:lhs} maybe rewritten by expanding the denominator as a binomial series
\begin{align}\nonumber
\int_{\mathrm{C}}\frac{f\left(z\right)(\log z-\pi\,i)}{\left(\omega^{2}+z^{2}\right)^{\frac{1}{2}}}\mathrm{d}z&=\int_{\mathrm{C}}\frac{f\left(z\right)(\log z-\pi\,i)}{z}\sum_{k=0}^{\infty}{{-\frac{1}{2}}\choose{k}}\left(\frac{\omega^{2}}{z^{2}}\right)^{k}\mathrm{d}z\\\nonumber
&=\sum_{k=0}^{\infty}{{-\frac{1}{2}\choose{k}}}\omega^{2k}\,\int_{C}\frac{f(z)\left(\log z-\pi i\right)}{z^{2k+1}}\mathrm{d}z\\
&=2\pi i\,\sum_{k=0}^{\infty}{{-\frac{1}{2}\choose{k}}}\omega^{2k}\,\bbint{0}{a}\frac{f(x)}{x^{2k+1}}\mathrm{d}x
\end{align}
Where we made use of the complex contour integral representation for the finite part \cite[eq.17]{tica}. 

The second term in the right hand side of equation \eqref{eq:lhs} is evaluated along the the contour $\mathrm{C'}$ but since only the paths $l_2$, $l_3$, $l_4$ and $l_5$ give non-vanishing contributions in the limit as $\epsilon\rightarrow 0$, we obtain
\begin{align}
\pi\,i\int_{\mathrm{C}'}\frac{f(z)}{\left(\omega^{2}+z^{2}\right)^{\frac{1}{2}}}\mathrm{d}z=-\,2\pi\,\int_{\epsilon}^{\omega-\epsilon}\frac{f(i\,y)}{\sqrt[]{\omega^{2}-y^{2}}}\mathrm{d}y - \,2\pi\,\int_{\epsilon}^{\omega-\epsilon}\frac{f(-i\,y)}{\sqrt[]{\omega^{2}-y^{2}}}\mathrm{d}y
\end{align}
Hence \eqref{eq:lhs} is given by
\begin{align}\nonumber\label{eq:LHS}
\int_{\mathrm{C}'}\frac{f\left(z\right)\log z}{\left(\omega^{2}+z^{2}\right)^{\frac{1}{2}}}\mathrm{d}z=2\pi i\,\sum_{k=0}^{\infty}{{-\frac{1}{2}\choose{k}}}\omega^{2k}&\bbint{0}{a}\frac{f(x)}{x^{2k+1}}\mathrm{d}x-\,2\pi\,\int_{\epsilon}^{\omega-\epsilon}\frac{f(i\,y)}{\sqrt[]{\omega^{2}-y^{2}}}\mathrm{d}y\\
&- \,2\pi\,\int_{\epsilon}^{\omega-\epsilon}\frac{f(-i\,y)}{\sqrt[]{\omega^{2}-y^{2}}}\mathrm{d}y
\end{align}
Combining the results \eqref{eq:LHS} with that of \eqref{eq:rhs} and taking the limit as $\epsilon\to 0$ on both sides
\begin{align}\nonumber\label{higt}
\int_{0}^{a}\frac{f(x)}{\sqrt[]{\omega^{2}+x^{2}}}\mathrm{d}x=\sum_{k=0}^{\infty}{{-\frac{1}{2}\choose{k}}}\omega^{2k}&\,\bbint{0}{a}\frac{f(x)}{x^{2k+1}}\mathrm{d}x-\frac{1}{\pi}\int_{0}^{\omega}\frac{\ln y\,(f\left(i\,y\right)+f\left(-i\,y\right))}{\sqrt[]{\omega^{2}-y^{2}}}\mathrm{d}y\\
&-\frac{1}{2i}\int_{0}^{\omega}\frac{f(i\,y)-f(-i\,y)}{\sqrt[]{\omega^{2}-y^{2}}}\mathrm{d}y
\end{align}
The final result is obtained by expanding  $f(x)$ in the last two integrals above as a Taylor series about the origin and interchanging the order of integration and summation. The validity of this step is ensured by the assumption that the series representation of $f(x)$ has an infinite radius of convergence. The following results given in \cite[p324, eq.2 and 3]{ryzhik} are then used
\begin{equation}
\int_{0}^{1}\frac{x^{2j+1}}{\sqrt{1-x^2}}\mathrm{d}x = \frac{(2j)!!}{(2j+1)!!},\,\,\int_{0}^{1}\frac{x^{2j}}{\sqrt{1-x^2}}\mathrm{d}x = \frac{\pi}{2}\frac{(2j)!!}{(2j+1)!!},
\end{equation}
and \cite[p538, eq.1]{ryzhik}
\begin{equation}
\int_{0}^{1}\frac{x^{2j}\ln x}{\sqrt{1-x^2}}\mathrm{d}x = \frac{(2j-1)!!}{(2j)!!}\frac{\pi}{2}\left(\sum_{k=1}^{2j}\frac{(-1)^{k-1}}{k}-\ln2\right).
\end{equation}
So that after a change of variable $y=\omega x$ and a few simplifications, we obtain
\begin{equation}
\int_{0}^{\omega}\frac{y^{2j+1}}{\sqrt{\omega^{2}-y^{2}}}\mathrm{d}y = \frac{\sqrt{\pi}\omega^{2j+1}j!}{2\Gamma\left(j+\frac{3}{2}\right)}
\end{equation}
and
\begin{equation}
\int_{0}^{\omega}\frac{y^{2j}\ln y}{\sqrt{\omega^{2}-y^{2}}}\mathrm{d}y = \frac{\sqrt{\pi}\Gamma\left(j+\frac{1}{2}\right)\,\omega^{2j}}{4j!}\left(H_{j-\frac{1}{2}}-H_j+2\ln\omega\right),
\end{equation}
where $j=0,1,\dots$ and $H_l$ is a harmonic number.

The proof is concluded by obtaining the condition for the convergence of the infinite series in the right hand side of equation \eqref{higt}. The steps are again identical to \cite[eq.63-64]{tica} so that the condition for convergence is $\omega<a$.
\end{proof}

\subsection{Example}
We now apply Theorem-\ref{orat} to obtain a representation of the integral
\begin{equation}\label{kyit}
\int_{0}^{\infty}\frac{e^{-a\left(x-b\right)^2}}{\sqrt{\omega^2+x^2}}\mathrm{d}x = e^{-ab^2}\int_{0}^{\infty}\frac{e^{-ax^2+2abx}}{\sqrt{\omega^2+x^2}}\mathrm{d}x,\qquad a>0
\end{equation}
that is suitable for computation for $\omega\ll 1$. This integral is relevant in the calculation of the effective index of refraction when a Gaussian wave packet is incident to a quantum well \cite{alvin}. The regime wherein $\omega\ll 1$ corresponds to the physical case when the well has a shallow depth.
If we let $a=\alpha$ and $2ab=\beta$ in equation \eqref{kyit}, then from Theorem-\ref{orat},
\begin{align}\nonumber\label{hinap}
\int_{0}^{\infty}\frac{e^{-\alpha x^2+\beta x}}{\sqrt{\omega^2+x^2}}\mathrm{d}x =&\sum_{j=0}^{\infty}{-\frac{1}{2}\choose{j}}\omega^{2j}\bbint{0}{\infty}\frac{e^{-\alpha x^2+\beta x}}{x^{2j+1}}\mathrm{d}x - \frac{\sqrt{\pi}}{2}\sum_{j=0}^{\infty}\frac{(-1)^{j}j!\,A_j\,(\beta\omega)^{2j+1}}{\Gamma\left(j+\frac{3}{2}\right)}\\
&-\frac{1}{2\sqrt{\pi}}\sum_{j=0}^{\infty}\frac{(-1)^{j}\Gamma\left(j+\frac{1}{2}\right)B_{j}\,(\beta\omega)^{2j}}{j!}\left(H_{j-\frac{1}{2}}-H_{j}+2\ln\omega\right)
\end{align}
where 
\begin{equation}
A_j = \frac{1}{(2j+1)!\beta^{2j+1}}\left.\frac{\mathrm{d}^{2j+1}}{\mathrm{d}x^{2j+1}}e^{-\alpha x^2+\beta x}\right|_{x=0} = \sum_{n=0}^{j}\frac{1}{(2j-2n+1)!\,n!}\left(-\frac{\alpha}{\beta^{2}}\right)^{n}
\end{equation}
and
\begin{equation}
B_j = \frac{1}{(2j)!\beta^{2j}}\left.\frac{\mathrm{d}^{2j}}{\mathrm{d}x^{2j}}e^{-\alpha x^2+\beta x}\right|_{x=0} = \sum_{n=0}^{j}\frac{1}{(2j-2n)!\,n!}\left(-\frac{\alpha}{\beta^{2}}\right)^{n}
\end{equation}
while the finite part integral is calculated in the Appendix and is given by
\begin{align}\nonumber
\bbint{0}{\infty}\frac{e^{-\alpha x^2+\beta x}}{x^{2j+1}}\mathrm{d}x =& \sum_{i=0}^{j}\frac{(-1)^{j-i+1}\alpha^{j-i}\beta^{2i}}{2(2i)!(j-i)!}\left(\ln\alpha-\psi(j-i+1)\right)\\
&+\frac{\pi}{2}\sum_{i=0}^{j-1}\frac{(-1)^{j-i}\alpha^{j-i-\frac{1}{2}}\beta^{2i+1}}{(2i+1)!\,\Gamma\left(j-i+\frac{1}{2}\right)} +\frac{\alpha^{j}}{2}\sum_{n=2j+1}^{\infty}\frac{\beta^{n}\Gamma\left(\frac{n}{2}-j\right)}{n! \sqrt{\alpha^{n}}}.
\end{align}
Hence from equation \eqref{hinap}, we obtain following leading-order behavior for $\omega\ll 1$,
\begin{align}\nonumber\label{witp}
\int_{0}^{\infty}\frac{e^{-\alpha x^2+\beta x}}{\sqrt{\omega^2+x^2}}\mathrm{d}x =& -\frac{1}{2}\left(\ln\left(\frac{\alpha\omega^{2}}{4}\right)+\gamma\right)+\frac{1}{2}\sum_{n=1}^{\infty}\frac{\Gamma\left(\frac{n}{2}\right)}{n!}\left(\frac{\beta}{\sqrt{\alpha}}\right)^{n}-\beta\omega\\
&
+\mathcal{O}\left(\omega^{2}\right)+\mathcal{O}\left(\omega^{2}\ln\omega\right).
\end{align}
From the result above, we see that  the dominant contribution to the value of the integral for $0<\omega\ll 1$ comes from the logarithmic term which is part of the correction term \eqref{kor} missed out by the naive term by term integration in equation \eqref{hinap}.
A detailed exposition on the physical consequences of the results summarized in equations \eqref{hinap} and \eqref{witp} will be given in \cite{alvin}.

A special case of equation \eqref{hinap} when $\beta = 0$ is given by,
\begin{align}\nonumber\label{otip}
\int_{0}^{\infty}\frac{e^{-\alpha x^2}}{\sqrt{\omega^{2}+x^{2}}}\mathrm{d}x = -\frac{\sqrt{\pi}}{2}\,\sum_{k=0}^{\infty}\frac{(-\alpha)^{k}\,\omega^{2k}}{k!^{2}\,\Gamma\left(\frac{1}{2}-k\right)}\left(\ln \alpha-\psi(k+1)\right)\hspace{2.3cm}\\
-\frac{1}{2\sqrt{\pi}}\sum_{j=0}^{\infty}\frac{\Gamma\left(j+\frac{1}{2}\right)\left(\sqrt{\alpha}\,\omega\right)^{2j}}{j!^{2}}\left(H_{j-\frac{1}{2}}-H_j+2\ln\omega\right).
\end{align}
The integral above is evaluated in \cite[Eq. 25, p367]{ryzhik} as,
\begin{equation}
\int_{0}^{\infty}\frac{e^{-\alpha x^2}}{\sqrt{\omega^{2}+x^{2}}}\mathrm{d}x = \frac{1}{2}\mathrm{exp}\left(\frac{\alpha\omega^{2}}{2}\right)K_{0}\left(\frac{\alpha\omega^{2}}{2}\right),
\end{equation}
where $K_0$ is the modified Bessel function of the second kind. Hence, combining the expansions in equation \eqref{otip} into a single summation, we obtain the following representation,
\begin{align}\label{owir}
K_{0}\left(x\right) = \frac{e^{-x}}{\sqrt{\pi}}\sum_{k=0}^{\infty}\frac{\Gamma\left(k+\frac{1}{2}\right)\,\left(2x\right)^{k}}{k!^2}\left(2\,\psi(k+1)-\psi\left(k+\frac{1}{2}\right)-\ln\left(2x\right)\right)
\end{align}
where we wrote $x=\alpha\omega^2/2>0$ and used the definition $H_{k}= \gamma+\psi(k+1)$ for the Harmonic number.
We thus obtain the following leading-order behavior for the modified Bessel function as $x\to 0$ 
\begin{equation}
K_{0}(x) = -\ln\left(\frac{x}{2}\right)\left(1+\mathcal{O}\left(x^2\right)\right)-\gamma\left(1+\mathcal{O}\left(x^2\right)\right),
\end{equation}
which is identical to the tabulated result \cite{besselk}. The exact representation \eqref{owir} as it stand, however is not tabulated in \cite{abramowitz,NIST} and does not appear to follow from the compendium of general representations for $K_{n}(x)$ for $n=0,1,\dots$ given in \cite{bessel}.

\section{Conclusion}\label{conclusion}
We obtained an exact representation of the non-integral order generalized Stieltjes transform $S_{\lambda}[f]=\int_0^{\infty} f(x) (\omega+x)^{-\lambda}\mathrm{d}x$ of function $f(x)$ with entire complex extension and real positive $\omega$ using the method of finite part integration. This representation allowed us to readily examine and extract the leading-order behavior of $S_{\lambda}[f]$ in the regime of $\omega\ll 1$. We then applied these results to generate known exact representations of special functions as well as their asymptotic behaviors in the relevant regimes. Moreover, our result gave the asymptotic behavior of the integral relevant in the computation of the effective index of refraction of a shallow one-dimensional quantum well for the case of an incident Gaussian wave packet. As a corollary to these calculations, we also obtained an exact representation of the modified Bessel function of the second kind $K_{0}(x)$ for $x>0$. The results of the present work complement those obtained for $S_{n}[f]$ for integer $n$ for a unified treatment of the generalized Stieltjes transform via finite part integration. Here and in the previous work, we hinted on how finite part integration lends itself to effectively solving problems in physical applications which are made formidable by having to treat divergent integrals as they arise in calculations. Thus, the investigation of several of these physical applications and related problems will serve to motivate further development on the range of applicability of finite part integration.

\section*{Acknowledgment}
The authors acknowledge the Office of the Chancellor of the University of the Philippines Diliman, through the Office of the Vice Chancellor for Research and Development, for funding support through the Outright Research Grant 171711 PNSE. C.D.T acknowledges the Department of Science and Technology-Science Education Institute for a scholarship grant under ASTHRDP-NSC.

\section*{Appendix}
We evaluate the finite part integral \cite[eq.37]{tica}
\begin{equation}\label{FPI1}
\bbint{0}{\infty} \frac{f(x)}{x^{m}} \, \mathrm{d}x = \lim_{a\rightarrow\infty} \left[c_{m-1} \ln a + \sum_{k=m}^{\infty} \frac{c_k a^{k-m+1}}{(k-m+1)}\right],
\end{equation}
for the function $f(x) = e^{-bx^{2}}$ whose Taylor series expansion about $x=0$ is given by,
	\begin{equation}
	e^{-b\,x^{2}}=\sum_{k=0}^{\infty}\frac{\left(-1\right)^{k}\,b^{k}\,x^{2k}}{k!},\qquad b>0
	\end{equation}
	The series expansion occurs in even powers of the argument so that we need to consider odd values of $m=2j-1$ where $c_{m-1}\neq 0$ separately from the even $m=2j$ where $c_{m-1} = 0$ for $j=1,2,\dots$. Lets us first consider the former case, 
	\begin{equation}
	\bbint{0}{\infty}\frac{e^{-b\,x^{2}}}{x^{2j-1}}\mathrm{d}x
	\end{equation}
	Equation \eqref{FPI1} gives,
	\begin{equation}\label{lop}
	\bbint{0}{\infty}\frac{e^{-b\,x^{2}}}{x^{2j-1}}\,\mathrm{d}x=\lim_{a\rightarrow\,\infty}\left(\frac{\left(-1\right)^{j-1}\,b^{j-1}}{\left(j-1\right)!}\ln a+\sum_{k=j}^{\infty}\frac{\left(-1\right)^{k}\,b^{k}\,a^{2k-2j+2}}{k!\left(2k-2j+2\right)}\right)
	\end{equation}
	we rewrite the infinite series in the RHS of the equation above as a hypergeometric function,
	\begin{equation}
\sum_{k=j}^{\infty}\frac{\left(-1\right)^{k}\,b^{k}\,a^{2k-2j+2}}{k!\left(2k-2j+2\right)}
	=\frac{\left(-1\right)^{j}\,a^{2}\,b^{j}}{2\,\left(j\right)!}{_2F_2}\left(1,\,1;\,j+1,\,2;\,-a^{2}\,b\right)
	\end{equation}
	and make use of its asymptotic expansion \cite[eq.40]{tica} as $a\to\infty$, this is given by
	\begin{equation}
	\lim_{a\to\infty}{_2F_2}\left(1,\,1;\,j+1,\,2;\,-a^{2}\,b\right) = \frac{j}{a^{2}\,b}\left(2\ln\,a+\ln\,b-\psi\left(j\right)\right)
	\end{equation}
	Hence equation \eqref{lop} assumes the form
\begin{equation}\label{huy}
	\bbint{0}{\infty}\frac{e^{-b\,x^{2}}}{x^{2j-1}}\mathrm{d}x = \frac{\left(-1\right)^{j}\,b^{j-1}}{2\,\left(j-1\right)!}\left(\ln\,b-\psi\left(j\right)\right),\,\,j=1,2,\dots,\,\,b>0.
	\end{equation}
	
	For even $m=2j$, $j=1,2,\dots$, we consider the finite part integral
	\begin{equation}
	\bbint{0}{\infty}\frac{e^{-b\,x^{2}}}{x^{2j}}\mathrm{d}x
	\end{equation}
	performing the same set of steps as above, we obtain
\begin{equation}\label{gik}
	\bbint{0}{\infty}\frac{e^{-b\,x^{2}}}{x^{2j}}\mathrm{d}x=\lim_{a\rightarrow\,\infty}\frac{\left(-1\right)^{j}\,a\,b^{j}}{j!}{_2F_2}\left(1,\,\frac{1}{2};\,j+1,\,\frac{3}{2};
	\,-a^{2}\,b\right)
	\end{equation}
	To evaluate the limit, we make use of the appropriate asymptotic expansion of the hypergeometric function as $a\rightarrow\,\infty$ for the case of simple pole which is given in \cite[eq.49]{tica}. This gives
	\begin{equation}\label{dti}
	\bbint{0}{\infty}\frac{e^{-b\,x^{2}}}{x^{2j}}\mathrm{d}x=\frac{\left(-1\right)^{j}\,b^{j-\frac{1}{2}}\,\pi}{2\,\Gamma\left(j+\frac{1}{2}\right)}.
	\end{equation}
    
We now obtain the finite part of the divergent integral
\begin{equation}
\int_{0}^{\infty}\frac{e^{-\alpha x^{2}+\beta x}}{x^{2j+1}}\mathrm{d}x,\qquad \alpha>0, j=0,1,2,\dots
\end{equation}
by replacement of the non-integrable origin by $\epsilon>0$ to temporarily remove the divergence, that is,
\begin{equation}
\int_{0}^{\infty}\frac{e^{-\alpha x^{2}+\beta x}}{x^{2j+1}}\mathrm{d}x\to\int_{\epsilon}^{\infty}\frac{e^{-\alpha x^{2}+\beta x}}{x^{2j+1}}\mathrm{d}x,\qquad\epsilon>0.
\end{equation}
We then 
substitute the following expansion.
\begin{equation}
e^{\beta x} = \sum_{n=0}^{2j}\frac{\beta^{n}}{n!}x^{n}+\sum_{n=2j+1}^{\infty}\frac{\beta^{n}}{n!}x^{n}
\end{equation}
and interchange the order of integration and summation and take the limit as $\epsilon\to 0$,
\begin{equation}\label{tiw}
\bbint{0}{\infty}\frac{e^{-\alpha x^{2}+\beta x}}{x^{2j+1}}\mathrm{d}x = \sum_{n=0}^{2j}\frac{\beta^{n}}{n!}\bbint{0}{\infty}\frac{e^{-\alpha x^{2}}}{x^{2j+1-n}}\mathrm{d}x+\sum_{n=2j+1}^{\infty}\frac{\beta^{n}}{n!}\int_{0}^{\infty}x^{n-2j-1}e^{-\alpha x^{2}}\mathrm{d}x.
\end{equation}
The integral in the second term of the equation above is convergent and evaluates to,
\begin{equation}\label{kiq}
\int_{0}^{\infty}x^{n-2j-1}e^{-\alpha x^{2}}\mathrm{d}x = \frac{\Gamma\left(\frac{n}{2}-j\right)}{2\alpha^{\frac{n}{2}-j}}.
\end{equation}
While in the first sum, we separate the terms corresponding to even $n =2i$ from odd $n=2i+1$, $i=0,1,2\dots$,
\begin{equation}\label{wop}
 \sum_{n=0}^{2j}\frac{\beta^{n}}{n!}\bbint{0}{\infty}\frac{e^{-\alpha x^{2}}}{x^{2j+1-n}}\mathrm{d}x =  \sum_{i=0}^{j}\frac{\beta^{2i}}{(2i)!}\bbint{0}{\infty}\frac{e^{-\alpha x^{2}}}{x^{2j-2i+1}}\mathrm{d}x+ \sum_{i=0}^{j-1}\frac{\beta^{2i+1}}{(2i+1)!}\bbint{0}{\infty}\frac{e^{-\alpha x^{2}}}{x^{2j-2i}}\mathrm{d}x
\end{equation}
Substituting equations \eqref{kiq} and \eqref{wop} into equation \eqref{tiw} and making use of the finite part integrals given in equations \eqref{huy} and \eqref{dti}, we finally obtain
\begin{align}\nonumber
\bbint{0}{\infty}\frac{e^{-\alpha x^2+\beta x}}{x^{2j+1}}\mathrm{d}x =& \sum_{i=0}^{j}\frac{(-1)^{j-i+1}\alpha^{j-i}\beta^{2i}}{2(2i)!(j-i)!}\left(\ln\alpha-\psi(j-i+1)\right)\\
&+\frac{\pi}{2}\sum_{i=0}^{j-1}\frac{(-1)^{j-i}\alpha^{j-i-\frac{1}{2}}\beta^{2i
+1}}{(2i+1)!\,\Gamma\left(j-i+\frac{1}{2}\right)} +\frac{\alpha^{j}}{2}\sum_{n=2j+1}^{\infty}\frac{\beta^{n}\Gamma\left(\frac{n}{2}-j\right)}{n!\sqrt{\alpha^{n}}}.
\end{align}
\end{document}